\documentclass[
  ,draft            % use draft while you are working on the paper
  ]
  {aipproc}

\layoutstyle{6x9}

\begin{document}

\title{Screening High-z GRBs with BAT Prompt Emission Properties}



\author{T. N. Ukwatta}{
  address={The George Washington University, Washington, D.C. 20052}
  ,altaddress={NASA Goddard Space Flight Center, Greenbelt, MD 20771}
}

\author{T. Sakamoto}{
  address={NASA Goddard Space Flight Center, Greenbelt, MD 20771}
  ,altaddress={The University of Maryland, Baltimore County, Baltimore, MD 21250}
%  ,altaddress={Oak Ridge Associated Universities, P.O. Box 117, Oak Ridge, Tennessee 37831-0117}
}

\author{M. Stamatikos}{
  address={NASA Goddard Space Flight Center, Greenbelt, MD 20771}
  ,altaddress={Oak Ridge Associated Universities, P.O. Box 117, Oak Ridge, Tennessee 37831-0117}
}

\author{N. Gehrels}{
  address={NASA Goddard Space Flight Center, Greenbelt, MD 20771}
}

\author{K. S. Dhuga}{
  address={The George Washington University, Washington, D.C. 20052}
}

%\author{K. S. Dhuga on behalf of the Swift-BAT team.}{
%  address={The George Washington University, Washington, D.C. 20052}
%}

\begin{abstract}
Detecting high-z GRBs is important for constraining the GRB
formation rate, and tracing the history of re-ionization and
metallicity of the universe. Based on the current sample of GRBs
detected by Swift with known redshifts, we investigated the
relationship between red-shift, and spectral and temporal
characteristics, using the BAT event-by-event data. We found
red-shift trends for the peak-flux-normalized temporal width T90,
the light curve variance, the peak flux, and the photon index in
simple power-law fit to the BAT event data. We have constructed
criteria for screening GRBs with high red-shifts. This will enable
us to provide a much faster alert to the GRB community of possible
high-z bursts.
\end{abstract}

\maketitle

%%%%%%%%%%%%%%%%%%%%%%%%%%%%%%%%%%%%%%%%%%%%
%% MAINMATTER
%%%%%%%%%%%%%%%%%%%%%%%%%%%%%%%%%%%%%%%%%%%%

\section{Introduction}
High-z GRBs offer the potential to probe the early universe into
the epoch of re-ionization. In Big Bang cosmology, the universe
went through a "dark age" when matter had cooled below $3000\
$Kelvin and became neutral. Re-ionization of matter occurred once
bodies such as quasars started radiating, initiating the
ionization of neutral matter, producing an ionized plasma. Bodies
we see now with a red-shift $20 > z > 6$ (150 million to one
billion years after the Big Bang)~\cite{lamb2000} emitted their
radiation in this re-ionization period. (The highest-z GRB
detected thus far is z=6.29~\cite{Tagliaferri2005}.)

High-z GRBs also have the potential to trace the star formation
rate and metallicity histories of the universe~\cite{lamb2000,
lamb2002}. Moreover, GRBs are 100-1000 times brighter at early
times than are high red-shift quasars. GRBs are expected to occur
out to z $>$ 10, whereas the distribution of quasars drops
significantly beyond z=3. Another benefit is that due to the
relatively clean neighborhoods of GRB progenitors, GRB afterglow
have simple power-law spectra with no emission lines. Thus GRBs
are clean probes of the intergalactic medium (IGM), whereas
quasars are contaminated by continuous material emission from the
central engine to the quasar's neighborhood. High-z GRB studies
will probe the IGM less than 1 Gyr after the Big Bang. In this
respect, the study of high-z GRBs provide a unique method to probe
the early universe.

Clearly, detecting high-z GRBs is very important for probing the
early universe. Currently, a significant number of GRBs with
redshifts around $z=2$ have been observed, but only a handful of
events for redshifts greater than $z=5.$ To help identify new GRBs
with high-z, we have developed an early-stage observational filter
on bursts, so that follow-up ground based observatories can be
notified, and they can subsequently attempt to get optical
red-shift measurements for these candidate high-z objects.

\section{Screening High-z Bursts using BAT DATA}
Our observational filter for candidate high-z GRBs is a set of
screening criteria, taking advantage of some prompt-emission
properties of GRBs which are analyzable within a few hours after
the burst.

\begin{figure}[htp]
  \includegraphics[height=.32\textheight]{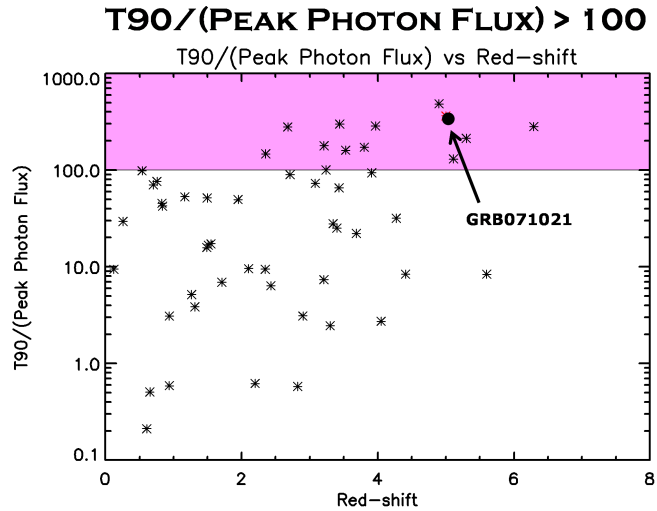}
  \caption{T90 normalized by the peak photon flux cut.\label{Fig01}}
\end{figure}

We have found trends between some observational characteristics in
the prompt emission from observed GRBs and their corresponding
redshifts ($z$):

1. The temporal period T90, normalized by the peak photon flux
(see Fig.~\ref{Fig01});

\begin{figure}[htp]
  \includegraphics[height=.32\textheight]{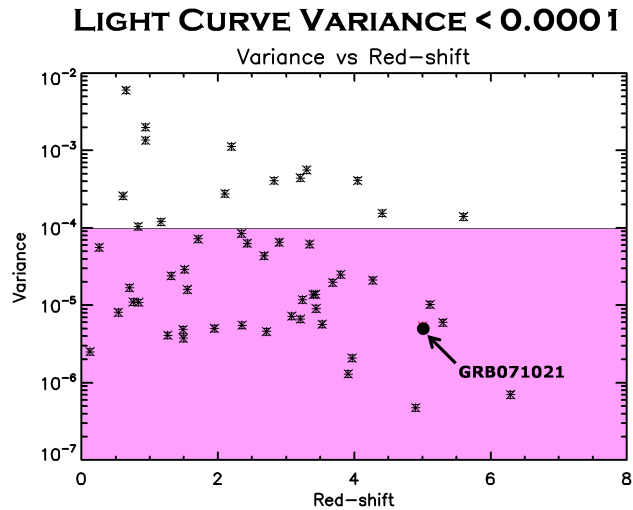}
  \caption{Normalized light curve variance cut.\label{Fig02}}
\end{figure}

2. The normalized light curve variance (using T100 duration) (see
Fig.~\ref{Fig02});

\begin{figure}[htp]
  \includegraphics[height=.32\textheight]{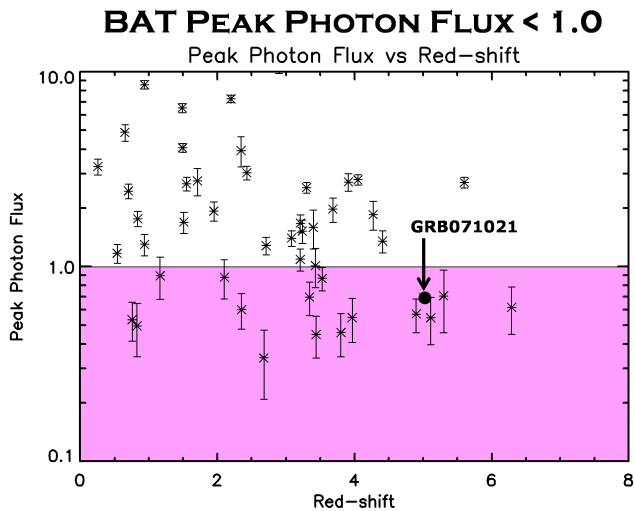}
  \caption{BAT peak photon flux cut.\label{Fig03}}
\end{figure}

3. The peak photon flux (Fig.~\ref{Fig03});

\begin{figure}[htp]
  \includegraphics[height=.32\textheight]{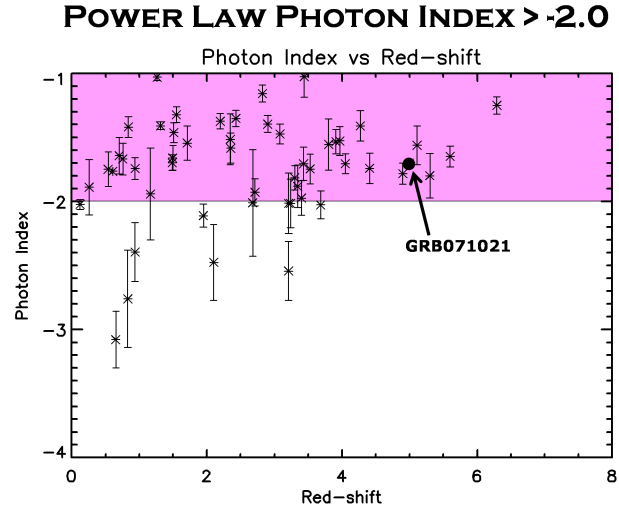}
  \caption{Photon Index cut.\label{Fig04}}
\end{figure}

4. The photon index derived from the simple power law fit to the
burst spectrum (Fig.~\ref{Fig04}).

Each characteristic is not sufficiently discriminatory to screen
for high-z GRBs by their own. However, the combined cuts on these
properties enable us to filter out high-z bursts with significant
probability.
\section{Discussion}
Using trends in pulse width, variance, peak flux, and photon power
law, we have formulated automated screening criteria to identify
possible high-z GRBs by utilizing the prompt emission properties
of GRBs within a few hours after the burst. This will enable more
intense ground-based observational campaigns for measuring
red-shifts of the detected GRBs.

With the limited available sample of GRBs having known high-z, the
probability of a GRB being selected by our criteria with a
redshift $\geq$ 5.0 is 45 \%, with a redshift $\geq$ 4.0 is 67 \%
and with a redshift $\geq$ 3.5 is 89 \% (Table~\ref{tab:a}). Based
on an analysis of Swift existing burst data, the criteria we used
will generate an alert once a month on average.

Currently, our screening criteria are implemented in automated
scripts and running in real time. Our filter for high-z has
already generated alerts on GRB 071021 (GCN 6967), GRB 071118 (GCN
7109) and GRB 071129 (GCN 7139), which enhanced ground based
follow-up observations.

We are working on implementing these screening criteria to BAT
TDRSS data. We also note that there is an ongoing effort to
formulate high-z screening criteria to include the XRT and UVOT
datasets.
\clearpage
\begin{table}
\begin{tabular}{lr}
\hline
    \tablehead{1}{l}{b}{Gamma-ray Burst Name }
  & \tablehead{1}{r}{b}{Measured Red-shift}   \\
\hline
GRB 050730 & 3.967\\
GRB 050814 & 5.3\\
GRB 050904 & 6.29\\
GRB 060115 & 3.53\\
GRB 060510B & 4.90\\
GRB 060522 & 5.11\\
GRB 060605 & 3.8\\
GRB 061110B & 3.44\\
GRB 070110 & 2.352\\
\hline
\end{tabular}
\caption{Selected GRBs using the high-z criteria from Swift GRBs
with known red-shifts.} \label{tab:a}
\end{table}
%\\ \\
%We will study the possibility of using results of the above
%studies to design a high-z early-alert system for Swift BAT
%TDRSS\footnote{Tracking and Data Relay Satellite System} data.
%TDRSS data comes down to earth in seconds after the GRB is
%detected and is used to measure the basic properties before the
%more reliable event by event data comes to the ground. The time
%gap between the GRB and the arrival time of event by event data
%can be a few hours. It is possible that during that gap, the GRB
%will already be unobservable. Thus, developing an early alert
%criteria using TDRSS data would be very useful in the study of
%high-z GRBs and consequently the early universe.

%%%%%%%%%%%%%%%%%%%%%%%%%%%%%%%%%%%%%%%%%%%%%%%%
%% BACKMATTER
%%%%%%%%%%%%%%%%%%%%%%%%%%%%%%%%%%%%%%%%%%%%%%%%

%%%%%%%%%%%%%%%%%%%%%%%%%%%%%%%%%%%%%%%%%%%%%%%%
%% The bibliography can be prepared using the BibTeX program or
%% manually.
%%
%% The code below assumes that BibTeX is used.  If the bibliography is
%% produced without BibTeX comment out the following lines and see the
%% aipguide.pdf for further information.
%%
%% For your convenience a manually coded example is appended
%% after the \end{document}
%%%%%%%%%%%%%%%%%%%%%%%%%%%%%%%%%%%%%%%%%%%%%%%%

%%%%%%%%%%%%%%%%%%%%%%%%%%%%%%%%%%%%%%%%%%%%%%%%
%% You may have to change the BibTeX style below, depending on your
%% setup or preferences.
%%
%%
%% For The AIP proceedings layouts use either
%%%%%%%%%%%%%%%%%%%%%%%%%%%%%%%%%%%%%%%%%%%%

\bibliographystyle{aipproc}   % if natbib is available
%\bibliographystyle{aipprocl} % if natbib is missing
%\bibliographystyle{unsrt}   % (uses file "plain.bst")

%%%%%%%%%%%%%%%%%%%%%%%%%%%%%%%%%%%%%%%%%%%
%% You probably want to use your own bibtex database here
%%%%%%%%%%%%%%%%%%%%%%%%%%%%%%%%%%%%%%%%%%%

%\bibliography{myrefs.bib}

%%%%%%%%%%%%%%%%%%%%%%%%%%%%%%%%%%%%%%%%%%%
%% Just a reminder that you may have to run bibtex
%% All of it up to \end{document} can be removed
%% if you don't like the warning.
%%%%%%%%%%%%%%%%%%%%%%%%%%%%%%%%%%%%%%%%%%%
%\IfFileExists{\jobname.bbl}{}
% {\typeout{}
%  \typeout{******************************************}
%  \typeout{** Please run "bibtex \jobname" to optain}
%  \typeout{** the bibliography and then re-run LaTeX}
%  \typeout{** twice to fix the references!}
%  \typeout{******************************************}
%  \typeout{}
% }

\end{document}